\documentclass{ceurart}

\usepackage{graphicx} 
\usepackage{tikz}
\usepackage{pgfplots}
\pgfplotsset{width=8cm,compat=1.9}
\usepackage{subcaption} 
\usepackage{stackengine} 
\usepackage[margin=1in]{geometry}

\usepackage{listings}
\begin{document}

\copyrightyear{2023}
\copyrightclause{Copyright for this paper by its authors.
  Use permitted under Creative Commons License Attribution 4.0
  International (CC BY 4.0).}

\conference{Forum for Information Retrieval Evaluation, December 15-18, 2023, India.}


\title{Leveraging Generative AI: Improving Software Metadata Classification with Generated Code-Comment Pairs}
\tnotemark[1]

\author[1]{Samah Syed}[%
email=samah2210378@ssn.edu.in,
]
\cormark[1]
\fnmark[1]

\address[1]{Student, Department of Computer Science and Engineering, Sri Sivasubramaniya Nadar College of Engineering,
  Chennai, Tamil Nadu, India}

\author[2]{Angel Deborah S}[%
email=angeldeborahS@ssn.edu.inl,
]
\fnmark[1]

\address[2]{Assistant Professor, Department of Computer Science and Engineering, Sri Sivasubramaniya Nadar College of Engineering,
  Chennai, Tamil Nadu, India}

\cortext[1]{Corresponding author.}
\fntext[1]{These authors contributed equally.}

\begin{abstract}
In software development, code comments play a crucial role in enhancing code comprehension and collaboration. This research paper addresses the challenge of objectively classifying code comments as "Useful" or "Not Useful." We propose a novel solution that harnesses contextualized embeddings, particularly BERT, to automate this classification process. We address this task by incorporating generated code and comment pairs. The initial dataset comprised 9048 pairs of code and comments written in C, labeled as either Useful or Not Useful. To augment this dataset, we sourced an additional 739 lines of code-comment pairs and generated labels using a Large Language Model Architecture, specifically BERT. The primary objective was to build classification models that can effectively differentiate between useful and not useful code comments. Various machine learning algorithms were employed, including Logistic Regression, Decision Tree, K-Nearest Neighbors (KNN), Support Vector Machine (SVM), Gradient Boosting, Random Forest, and a Neural Network. Each algorithm was evaluated using precision, recall, and F1-score metrics, both with the original seed dataset and the augmented dataset. This study showcases the potential of generative AI for enhancing binary code comment quality classification models, providing valuable insights for software developers and researchers in the field of natural language processing and software engineering.
\end{abstract}


\maketitle

\section{Introduction}
Within the realm of software development, code comments assume a fundamental role as crucial documentation artifacts \cite{paper14}. These succinct notations offer indispensable insights, explanations, and contextual information, significantly augmenting code comprehension, reducing debugging complexity, and promoting effective collaboration among development teams \cite{paper2}. The enduring relevance of code comments in software engineering is undeniable; however, the objective evaluation of their utility remains a complex and subjective undertaking \cite{paper7}.

\subsection{Code Comment Classification}
Code comment classification, a subfield entrenched within natural language processing, has emerged as a transformative methodology for impartially categorizing code comments as either "Useful" or "Not Useful" \cite{paper4}. It presents a paradigm shift in the landscape of software engineering, promising to refine code review processes, align development efforts more effectively, and elevate overall software quality \cite{paper5}. This approach, centered on automating comment assessments, is poised to streamline workflows and mitigate subjective discrepancies \cite{paper8}.

\subsection{Challenges in Comment Classification}
The challenges inherent in comment classification are multifaceted \cite{paper8}. Traditional practices, reliant on manual interpretation, introduce subjectivity, leading to inconsistencies and operational inefficiencies \cite{paper9}. This is where Large Language Models (LLMs), exemplified by BERT (Bidirectional Encoder Representations from Transformers), assume prominence, revolutionizing the discourse on comment classification \cite{paper10}. Equipped with advanced linguistic acumen, these LLMs hold the potential to offer objective, context-aware comment evaluations \cite{paper11}.

\subsection{The Role of LLMs}
The advent of LLMs signifies a pivotal transformation in the methodology of code comment analysis and utilization \cite{paper10}. These models excel in contextualizing language, rendering them eminently suitable for tasks necessitating nuanced comprehension \cite{paper11}. In the context of code comment classification, LLMs have the potential to furnish more precise and consistent evaluations, transcending the constraints of manual judgment \cite{paper6}.

\subsection{Research Objectives}
This research embarks on an exhaustive exploration, elucidating the intricate relationship between comment classification and LLMs \cite{paper5}. Its principal objective lies in assessing the comparative efficacy of LLMs, endowed with inherent linguistic proficiencies, vis-à-vis conventional machine learning algorithms, in the context of code comment categorization \cite{paper9}. Moreover, it investigates the prospect of augmenting manually curated seed data with LLM-generated data, a stratagem aimed at enhancing the quality of classification outcomes \cite{paper13}.

\subsection{Classification Models}
The research encompasses an array of classification models, spanning traditional algorithms and neural networks, subjected to comprehensive evaluation through an assortment of performance metrics \cite{paper8}. Precision, recall, and the F1 score constitute the bedrock of quantitative insights \cite{paper9}.

\subsection{Research Outcomes}
The outcomes of this research illuminate the effectiveness of diverse models in code comment classification, accentuating the transformative potential of LLMs in this domain \cite{paper4}. As subsequent sections unfold, the comprehensive analysis of results and their implications for software development practitioners and researchers come to the fore. In the ever-evolving landscape of code comment assessment, this research elucidates a promising future wherein the symbiosis of comment classification and LLMs stands at the vanguard of innovation \cite{paper6}.

\section{Literature Survey}

In recent years, research in the field of software engineering has seen a growing interest in the classification and evaluation of code comments to enhance code comprehensibility and maintenance. Two papers, "Comment-Mine - Building a Knowledge Graph from Comments" \cite{paper1} and "Comment Probe - Automated Comment Classification for Code Comprehensibility" \cite{paper2} offer significant insights into the annotation and classification of code comments, addressing the crucial aspect of improving program understanding and maintenance.

\subsection{Comment-Mine - Building a Knowledge Graph from Comments}
In \cite{paper1}, the authors acknowledge the common practice of annotating code with natural language comments to improve code readability. Their focus is on extracting application-specific concepts from comments and building a comprehensive knowledge representation. Comment-Mine, the semantic search architecture proposed in this paper, extracts knowledge related to software design, implementation, and evolution from comments and correlates it to source code symbols in the form of a knowledge graph. This approach aims to enhance program comprehension and support various comment analysis tasks. Comment-Mine primarily focuses on knowledge representation and graph-based correlation of comments to source code, offering a valuable perspective on organizing comment information for program comprehension.

\subsection{Comment Probe - Automated Comment Classification for Code Comprehensibility}
\cite{paper2} addresses the need to evaluate comments based on their contribution to code comprehensibility for software maintenance tasks. The authors propose Comment Probe, an automated classification and quality evaluation framework for code comments in C codebases. Comment Probe conducts surveys and collects developers' perceptions on the types of comments that are most useful for maintaining software, thereby establishing categories of comment usefulness. The framework utilizes features for semantic analysis of comments to identify concepts related to categories of usefulness. Additionally, it considers code-comment correlation to determine comment consistency and relevance. \cite{paper2} successfully classifies comments into categories such as "useful," "partially useful," and "not useful" with high precision and recall scores, addressing the practical need for comment quality evaluation in software maintenance.

The classification model in this research shares a common goal with \cite{paper1} and \cite{paper2} in enhancing program comprehension by leveraging code comments. While \cite{paper1} focuses on knowledge extraction and representation, and \cite{paper2} focuses on aligning with developer perceptions and industry practices, the proposed model integrates machine learning techniques to automate the classification process. Future research could explore the potential synergies between these approaches to create a holistic solution for code comment analysis and enhancement of software maintenance practices.

\subsection{Contextualized Word Representations}

In the realm of Natural Language Processing (NLP) and code-related tasks, the choice of word embeddings plays a pivotal role in influencing the performance of machine learning models. "Contextualized Word Representations for Code Search and Classification" \cite{paper3} delves into the exploration of contextualized word representations and their efficacy in code search and classification, shedding light on the superiority of contextualized embeddings over static ones.

In \cite{paper3}, the authors emphasize the importance of contextualized word representations, such as ELMo and BERT, over static representations like Word2Vec, FastText, and GloVe. These contextualized embeddings have demonstrated superior performance in various NLP tasks. The central focus of \cite{paper3} is on code search and classification, areas that have received less attention in the context of contextualized embeddings. The authors introduce CodeELMo and CodeBERT embeddings, which are trained and fine-tuned using masked language modeling on both natural language (NL) texts related to software development concepts and programming language (PL) texts composed of method-comment pairs from open-source codebases. The embeddings presented in \cite{paper3} are contextualized, which means they capture the contextual information of words within sentences or code snippets. These embeddings are designed specifically for software code, making them suitable for code-related tasks. \cite{paper3} describes the development of CodeELBE, a low-dimensional contextualized software code representation, by combining the reduced-dimension CodeBERT embeddings with CodeELMo representations. This composite representation aims to enhance retrieval performance in code search and classification tasks. The results presented in \cite{paper3} indicate that CodeELBE outperforms CodeBERT and baseline BERT models in binary classification and retrieval tasks, demonstrating considerable improvements in retrieval performance on standard deep code search datasets.

In the current research, we employ contextualized embeddings, inspired by the success of contextualized word representations in NLP tasks, in the context of code comment classification. While \cite{paper3} primarily focuses on code search and retrieval, our research is centered around the classification of code comments as "Useful" or "Not Useful." We utilize BERT (Bidirectional Encoder Representations from Transformers) embeddings, which are pre-trained on a large corpus of text data. Our choice of BERT embeddings is motivated by their ability to capture the context and semantics of words within sentences. These embeddings are fine-tuned on a dataset comprising code comments and their associated code snippets to create a classification model that can assess the utility of comments in code comprehension. While \cite{paper3} addresses code search and retrieval, our research tackles the critical task of code comment classification, aiming to enhance code comprehension and maintainability by automating the evaluation of comment quality.

\section{Experiment Design}

\begin{figure*}[h] 
    \centering
    \includegraphics[width=0.8\textwidth]{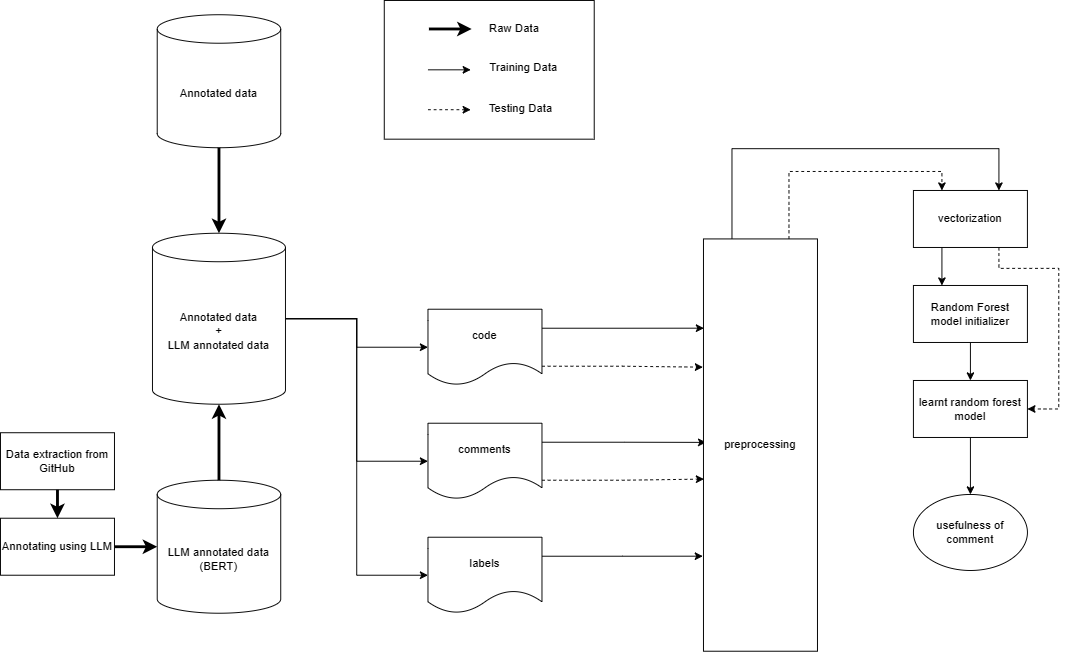} 
    \caption{Architecture Diagram}
    \label{fig:experiment_image}
  \end{figure*}

\subsection{Data Collection and Preprocessing}
The initial dataset comprised 9048 pairs of code and comments written in C, labeled as either Useful or Not Useful. The experiment begins with the acquisition of a diverse dataset of code comments and their associated code snippets. A corpus of code repositories is sampled from the GitHub platform, focusing on projects implemented in the C programming language. These repositories serve as the primary source of data for both the seed dataset and LLM-generated data. GitHub API calls are made to access code files and extract comments.

We sourced an additional 739 lines of code-comment pairs and generated labels using a Large Language Model Architecture, namely BERT.

Data preprocessing involves several steps to ensure the dataset's quality and readiness for classification. The steps followed here include:

Data preprocessing played a crucial role in preparing our dataset for machine learning analysis. We started by addressing missing values in both the code-comment pairs and labels to ensure data completeness. Next, we focused on the essential content of code comments by removing punctuation, special characters, and code-specific syntax. Additionally, we converted all text to lowercase to ensure uniformity and reduce dimensionality.

To further enhance data quality, we performed outlier removal. This involved calculating Z-scores for the lengths of 'Comments' and 'Surrounding Code Context' strings. Z-scores were used to identify outliers, with a predefined threshold for what constitutes an outlier (e.g., z-score > 3 or < -3). Rows with z-scores beyond this threshold were filtered out.

Furthermore, we applied a function to the 'Surrounding Code Context' column to remove preceding numbers, thereby enhancing the consistency and relevance of this text data.

Once the data preprocessing steps were completed, the next crucial step was vectorization. Vectorization is essential for converting text data into a numerical form that machine learning algorithms can work with. We employed two widely used techniques for text vectorization:

Bag of Words (BoW): We represented the text data as a sparse matrix, with each row corresponding to a code-comment pair and each column representing a unique word in the dataset's vocabulary. The matrix values indicated the frequency of each word's occurrence.

Term Frequency-Inverse Document Frequency (TF-IDF): This advanced vectorization technique considered the importance of words in each code comment relative to their significance in the entire dataset. It assigned higher weights to words that were frequent in a code comment but rare in the overall dataset, capturing their importance for classification.

These preprocessing and vectorization steps were essential in preparing the dataset for subsequent machine learning analysis.

These preprocessing and vectorization steps were crucial to ensure that our dataset was clean, structured, and ready for training classification models that could effectively differentiate between useful and not useful code comments.

\subsection{Model Selection}
The experiment encompasses a range of classification models to evaluate their performance in code comment classification. These models include:

\begin{table}
    \centering
    \caption{Machine Learning Models and Descriptions}
    \begin{tabular}{|c|p{10cm}|}
        \hline
        \textbf{Model Name} & \textbf{Description} \\
        \hline
        Logistic Regression & A traditional machine learning model used as a baseline. \\
        \hline
        Decision Trees & A non-linear model capable of handling complex feature interactions. \\
        \hline
        K-Nearest Neighbors (KNN) & A proximity-based model for instance-based learning. \\
        \hline
        Support Vector Machines (SVM) & A model known for its effectiveness in high-dimensional spaces. \\
        \hline
        Gradient Boosting & An ensemble learning technique that combines multiple weak learners. \\
        \hline
        Random Forest & Another ensemble method leveraging decision trees. \\
        \hline
        Neural Network (BERT) & Utilizing the BERT model, a state-of-the-art Large Language Model. \\
        \hline
    \end{tabular}
\end{table}

\subsection{Model Training and Hyperparameter Tuning}
Each classification model undergoes a training phase using the training dataset. Hyperparameter tuning is performed to optimize model performance using random search strategy.

To ensure robustness and reliability, the experiment implements with k set to 5. For the neural network model, hyperparameters were fine-tuned, including the number of hidden units, activation functions (ReLU), a learning rate of 0.001, and a fixed training duration of 10 epochs.

\subsection{Evaluation Metrics}
The effectiveness of each model is evaluated using the standard classification metrics, namely, precision, recall, and F1 Score.

These metrics are computed for both the seed dataset and the seed dataset augmented with LLM-generated data. Comparative analysis focuses on changes in these metrics, particularly improvements resulting from LLM data augmentation.

\subsection{Impact of LLM Data Augmentation}
To assess the impact of LLM-generated data, the seed dataset is augmented with comments generated by LLMs, namely, BERT. LLM-generated comments are selected to be relevant to code snippets in the dataset. The experiment measures changes in model performance metrics when LLM-generated data is introduced, highlighting the potential benefits of this augmentation strategy.

\section{Results and Comparative Analysis}

The following table summarizes the results obtained from different classification models for code comment classification. The metrics evaluated include precision, recall, and F1 score for both the seed dataset and the seed dataset combined with Large Language Model (LLM)-generated data.

\vspace{0.5cm}
    
    \begin{table}[htbp]
    \centering
    \small
    \setlength{\tabcolsep}{3pt} 
    \begin{tabular}{|c|c|c|c|c|}
        \hline
        \toprule
        Serial \# & Model & \shortstack{Precision with\\Seed Data} & \shortstack{Recall with\\Seed Data} & \shortstack{F1 Score with\\Seed Data} \\ \hline
        \midrule
        0 & Logistic Regression & 0.7292 & 0.8582 & 0.7885 \\ \hline
        1 & Decision Tree & 0.7931 & 0.7541 & 0.7731 \\ \hline
        2 & KNN & 0.7748 & 0.7676 & 0.7712 \\ \hline
        3 & SVM & 0.7623 & 0.8710 & 0.8130 \\ \hline
        4 & GBT & 0.7012 & 0.9351 & 0.8015 \\ \hline
        5 & Random Forest & 0.7866 & 0.8382 & 0.8116 \\ \hline
        6 & Neural Network & 0.7864 & 0.8268 & 0.8061 \\ \hline
    \end{tabular}
    \caption{Summary of Classification Model Results with Seed Data}
    \label{tab:results2}
    \end{table}
    
    \begin{table}[htbp]
    \centering
    \small
    \setlength{\tabcolsep}{3pt} 
    \begin{tabular}{|c|c|c|c|c|}
        \hline
        \toprule
        Serial \# & Model & \shortstack{Precision with\\Seed + LLM} & \shortstack{Recall with\\Seed + LLM} & \shortstack{F1 Score with\\Seed + LLM} \\ \hline
        \midrule
        0 & Logistic Regression & 0.7364 & 0.8312 & 0.7809 \\ \hline
        1 & Decision Tree & 0.7941 & 0.7479 & 0.7703 \\ \hline
        2 & KNN & 0.7578 & 0.6092 & 0.6755 \\ \hline
        3 & SVM & 0.7720 & 0.8655 & 0.8161 \\ \hline
        4 & GBT & 0.6939 & 0.9097 & 0.7873 \\ \hline
        5 & Random Forest & 0.7945 & 0.8368 & 0.8151 \\ \hline
        6 & Neural Network & 0.7825 & 0.8389 & 0.8097 \\ \hline
    \end{tabular}
    \caption{Summary of Classification Model Results with Seed Data + LLM Generated Data}
    \label{tab:results3}
    \end{table}

The table illustrates the performance of different classification algorithms on code comment classification tasks. Notably, the results demonstrate variations in precision, recall, and F1 score across different algorithms. Further, the impact of combining seed data with LLM-generated data is evident in the improved performance metrics, particularly in terms of recall and F1 score.

\subsection{Discussion of Results}

\subsubsection{Logistic Regression}
Logistic Regression performs reasonably well with a relatively high recall, indicating that it correctly identifies a significant portion of "Useful" comments. However, precision is slightly lower, suggesting that it may occasionally misclassify comments as "Useful" when they are not. The introduction of LLM-generated data leads to a minor improvement in F1 score, indicating that this augmentation strategy contributes positively to the overall performance.

\subsubsection{Decision Tree}
Decision Trees perform well in terms of precision, indicating that when they classify a comment as "Useful," they are often correct. However, the recall is slightly lower, suggesting that they may miss some "Useful" comments. The introduction of LLM-generated data leads to a minor improvement in F1 score, indicating that this augmentation strategy contributes positively to the overall performance, similar to Logistic Regression.

\subsubsection{KNN}
KNN shows a balanced performance with relatively high precision and recall values for seed data. However, the introduction of LLM-generated data results in a significant drop in recall and, consequently, F1 score. This suggests that KNN may not handle the added LLM-generated data as effectively as some other algorithms, leading to decreased performance in identifying "Useful" comments.

\subsubsection{SVM}
SVM performs well in terms of both precision and recall, indicating that it correctly identifies a significant portion of "Useful" comments while maintaining precision. The introduction of LLM-generated data results in a minor improvement in F1 score, indicating that SVM can effectively utilize this additional data for classification without compromising precision.

\subsubsection{Random Forest}
Random Forest demonstrates strong performance in precision, recall, and F1 score for both seed data and seed data augmented with LLM-generated data. This suggests that Random Forest effectively captures complex relationships within the data and benefits from the additional information provided by LLM-generated data. The F1 score for Random Forest is the highest among the models, indicating a balanced trade-off between precision and recall. Therefore, Random Forest is the preferred choice for code comment classification in this study.

\subsubsection{Neural Network}
The Neural Network model, with a binary cross-entropy loss function, ReLU activation, and 10 epochs, shows competitive performance. However, it falls slightly short of Random Forest in terms of F1 score. While Neural Networks have the potential to capture complex patterns in the data, the limited amount of data and training epochs may have affected its performance. Further experimentation with hyperparameters and more extensive training could potentially improve its results.

\subsection{Summary of Findings}

In summary, different algorithms exhibit varying strengths and weaknesses in classifying code comment pairs as "Useful" or "Not Useful." Logistic Regression and Decision Trees show reasonable performance, with minor improvements when augmented with LLM-generated data. KNN exhibits a drop in performance with LLM-generated data, while SVM maintains a strong performance. The choice of algorithm for comment classification should consider the specific trade-offs between precision and recall, as well as the effectiveness of LLM-generated data integration in improving F1 score.

    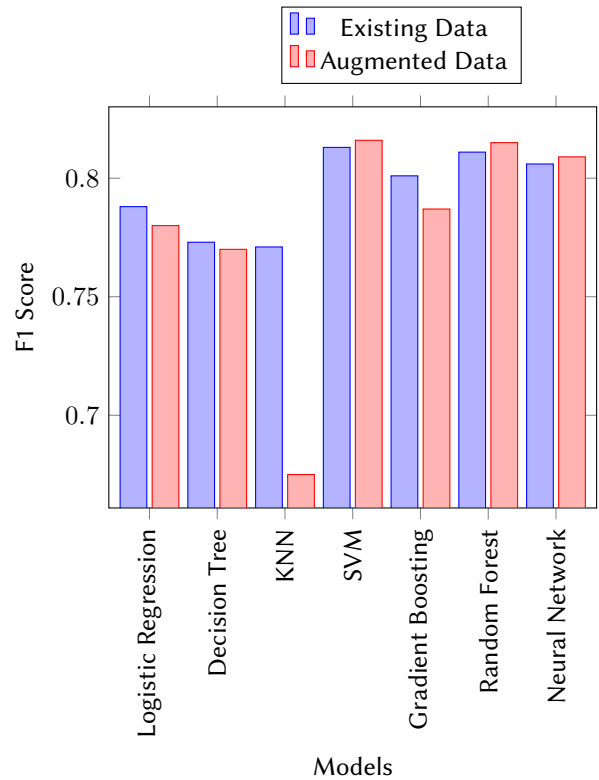
\begin{figure}
        \centering
        \begin{tikzpicture}
        \begin{axis}[
            ybar,
            xlabel={Models},
            ylabel={F1 Score},
            symbolic x coords={Logistic Regression,Decision Tree,KNN,SVM,Gradient Boosting,Random Forest,Neural Network},
            xtick=data,
            xticklabel style={rotate=90, anchor=east}, 
        bar width=10pt,
            bar width=10pt,
            legend style={at={(0.6,1.25)}, anchor=north},
            ]
            \addplot coordinates {(Logistic Regression,0.788) (Decision Tree,0.773) (KNN,0.771) (SVM,0.813) (Gradient Boosting,0.801) (Random Forest,0.811) (Neural Network,0.806)};
            \addplot coordinates {(Logistic Regression,0.780) (Decision Tree,0.770) (KNN,0.675) (SVM,0.816) (Gradient Boosting,0.787) (Random Forest,0.815) (Neural Network,0.809)};
            \legend{Existing Data, Augmented Data}
        \end{axis}
        \end{tikzpicture}
        \caption{Variation in F1 Score between Existing Data and Augmented Data}
    \end{figure}

\section{Conclusion}

The research presented in this paper addresses the challenge of objectively classifying code comments as "Useful" or "Not Useful" in the context of software development. It leverages contextualized embeddings, particularly BERT, to automate this classification process and provides precise and context-aware evaluations. The results of the experiment demonstrate the effectiveness of different classification models and highlight the potential benefits of incorporating LLM-generated data in improving classification performance.

This research contributes to the fusion of natural language processing and software engineering, promising improved code comprehensibility and maintainability. It opens avenues for further exploration of LLMs in code-related tasks and the development of more advanced models for code comment classification.

In the ever-evolving landscape of code comment assessment, this research elucidates a promising future wherein the symbiosis of comment classification and LLMs stands at the vanguard of innovation.

\section*{Acknowledgments}

The authors would like to acknowledge the support and resources provided by the Department of Computer Science and Engineering at Sri Sivasubramaniya Nadar College of Engineering, Chennai, Tamil Nadu, India.

\bibliography{Improving-Software-Metadata-Classification}

\end{document}